\documentclass{aa}  

\usepackage{xcolor}

\usepackage{ulem}

\usepackage{upgreek}
\usepackage{float}
\usepackage{textcase}
\usepackage{graphicx}
\usepackage{xcolor}
\usepackage{txfonts}
\usepackage{lipsum}
\usepackage{amsmath}
\usepackage{subcaption}         % necessary for continued figures, example in section 3
\usepackage{lscape}             % to rotate a single page table, example in appendix.
\usepackage{placeins}           % useful with \FloatBarrier, to keep 
\usepackage{natbib}
\usepackage{tabularx}
\usepackage{makecell}
\usepackage{geometry} % 调整页边距
\usepackage{booktabs} % 用于制作更专业的三线表
\usepackage{array} % 用于调整列格式
\usepackage{amsmath} % 用于数学模式（如像素尺度中的上标）
\usepackage[colorlinks=true,
            linkcolor=blue,
            citecolor=blue,
            urlcolor=blue,
            bookmarks=true,
            bookmarksopen=true]{hyperref}
\bibpunct{(}{)}{;}{a}{}{,}              
\begin{document}

%%%%%%%%%%%%%%%%%%%%%%%%%%%%%%%%%%%%%%%%
% if you use custom commands in your title,
% ensure to check your title when submitting!
%%%%%%%%%%%%%%%%%%%%%%%%%%%%%%%%%%%%%%%%
    \title{SN\,2024abfl: A Low-Luminosity Type IIP Supernova in NGC\,2146 from a Low-Mass Red Supergiant Progenitor}

    %\subtitle{Subtitle}

%%%%%%%%%%%%%%%%%%%%%%%%%%%%%%%%%%%%%%%%
% Please separate each author with the \and command
%
% Please do not include ORCIDs next to author names.
% Only ORCIDs authenticated by individual authors in EDPS
% editorial system will be taken into account.
% ORCIDs included here will be removed.
%%%%%%%%%%%%%%%%%%%%%%%%%%%%%%%%%%%%%%%%

    \author{Xiaohan Chen\inst{1,2}
        \and Ning-Chen Sun\inst{3,1,4}\fnmsep\thanks{Email: sunnc@ucas.ac.cn}
        % assistance in data processing
        \and Qiang Xi\inst{3,1}
        \and Chun Chen\inst{5,6,7}
        \and Yu Zhang\inst{1} 
        \and Zexi Niu\inst{3,1}
        % observation schedule (Xinglong)
        \and Junjie Jin\inst{1} 
        \and Yiming Mao\inst{1,3}
        \and Beichuan Wang\inst{1,3}
        % observation schedule (LT, TRT, NOT)
        \and Samaporn Tinyanont\inst{8}
        \and Krittapas Chanchaiworawit\inst{8} 
        \and Kanthanakorn Noysena\inst{8}
        \and David Aguado\inst{9,10}
        \and Ismael Perez-Fournon\inst{9,10}
        \and Fr\'ed\'erick Poidevin\inst{9,10}
        \and Justyn R. Maund\inst{13}
        % assistance in comments and observation
        \and Xunhao Chen\inst{1,3}
        \and Pengliang Du\inst{1}
        \and David López Fernández-Nespral\inst{9,10}
        \and Liguo Fang\inst{1}
        \and Guolin Gao\inst{1,2}
        \and Jiupeng Guo\inst{1}
        \and Min He\inst{1}
        \and Xinyi Hong\inst{1,3}
        \and Zhigang Hou\inst{1}
        \and Qingzheng Li\inst{1}
        \and Wenxiong Li\inst{1}
        \and Tongyu Liu\inst{1}
        \and Alicia López-Oramas\inst{9,10}
        \and Haiyang Mu\inst{1}
        \and César Rojas-Bravo\inst{3,1}
        \and Jianfeng Tian\inst{1}
        \and Jinhu Wang\inst{1}
        \and Lingzhi Wang\inst{15,16}
        \and Rui Wang\inst{1}
        \and Yanan Wang\inst{1}
        \and Ziyang Wang\inst{11,12}
        \and Klaas Wiersema\inst{14}
        \and Ying Wu\inst{1}
        \and Guo Zhen\inst{17,18}
        \and Jie Zheng\inst{1}
        \and Guoyin Zhu\inst{1}
        \and Yinan Zhu\inst{1}
        % corresponding authors
        \and Zhou Fan\inst{1,3}\fnmsep\thanks{Email: zfan@nao.cas.cn}
        \and Jing Li\inst{2}\fnmsep\thanks{Email: lijing@bao.ac.cn}
        \and Hong Wu\inst{1,3}
        \and Jifeng Liu\inst{1,3,4}}

    \institute{National Astronomical Observatories, Chinese Academy of Sciences, Beijing 100101, China
                %\email{sunnc@ucas.ac.cn}
                %\thanks{Shows the usage of elements in the institute field}
            \and School of Physics and Astronomy, China West Normal University, Nanchong 637002, China
            \and School of Astronomy and Space Science, University of Chinese Academy of Sciences, Beijing 100049, China
            \and Institute for Frontiers in Astronomy and Astrophysics, Beijing Normal University, Beijing, 102206, China
            \and School of Physics and Astronomy, Sun Yat-sen University, Zhuhai 519082, China
            \and CSST Science Center for the Guangdong-Hong Kong-Macau Greater Bay Area, Sun Yat-sen University, Zhuhai 519082, China
            \and Dipartimento di Fisica, Universit\`a di Napoli ``Federico II'', Compl. Univ. di Monte S. Angelo, Via Cinthia, I-80126, Napoli, Italy
            \and National Astronomical Research Institute of Thailand, 260 Moo 4, Donkaew, Maerim, Chiang Mai, 50180, Thailand
            \and Instituto de Astrofísica de Canarias, Vía L\'actea, 38205 La Laguna, Tenerife, Spain
            \and Universidad de La Laguna, Departamento de Astrofísica, 38206 La Laguna, Tenerife, Spain
            \and School of Physics and Astronomy, Beijing Normal University, Beijing 100875, China 
            \and Department of Physics, Faculty of Arts and Sciences, Beijing Normal University, Zhuhai 519087, China
            \and Department of Physics, Royal Holloway, University of London, Egham, TW20 0EX, United Kingdom
            \and Centre for Astrophysics Research, University of Hertfordshire, Hatfield, AL10 9AB, UK
            \and Chinese Academy of Sciences South America Center for Astronomy (CASSACA), National Astronomical Observatories, CAS, Beijing
            100101, China
            \and Departamento de Astronomía, Universidad de Chile, Las Condes, 7591245 Santiago, Chile
            \and Instituto de Física y Astronomía, Universidad de Valparaíso, ave. Gran Bretaña, 1111, Casilla 5030, Valparaíso, Chile 
            \and Millennium Institute of Astrophysics, Nuncio Monseñor Sotero Sanz 100, Of. 104, Providencia, Santiago, Chile 
            \\}
            
    \date{Received September 30, 20XX}

% \abstract{}{}{}{}{}
% 5 {} token are mandatory
    
    \abstract 
    {Type IIP supernovae (SNe IIP) exhibit a significant diversity in their explosion properties, yet the physical mechanisms driving this diversity remain unknown. 
    In this work, we present photometric and spectroscopic observations of SN\,2024abfl, a SN~IIP in NGC\,2146 with a directly detected red supergiant (RSG) progenitor.
    %covering the ultraviolet-optical bands from t = 4.4 to t = 379.0 days, spectroscopic observations from t = 3.6 to 198.6 days, and a comprehensive analysis of SN\,2024abfl. 
    We find it has a low plateau luminosity ($M_{V}\sim-15$ mag) and a relatively long plateau length ($\sim126.5$ days).  
    By fitting a semi-analytical model, we estimated a 
    $^{56}$Ni mass of $\sim0.009$ $M_\odot$, 
    an initial kinetic energy of $\sim0.42$ foe, 
    an initial thermal energy of $\sim0.03$ foe
    %, a progenitor radius of $\sim9.3\times10^{13}$ cm 
    and an ejecta mass of $\sim8.3$ $M_\odot$. 
    %The $^{56}$Ni mass is consistent with the value derived from the tail bolometric luminosity independently, which is $\sim0.027$ $M_\odot$. 
    The spectral evolution of SN\,2024abfl is similar to those of other SNe IIP, except for much lower ejecta velocities at similar epochs. At later epochs, we find a relatively high-velocity H$\alpha$ absorption feature at $\sim-4000$ km s$^{-1}$, possibly due to a fast-moving plume of matter in the inner ejecta, and two emission features at $\pm2000$ km s$^{-1}$, possibly caused by CSM interaction. %show up at $\pm2000$ km s$^{-1}$, possibly caused by CSM interaction. 
    %The SN enters the nebular phase from $t=138.0$ days and 
    We estimate the progenitor mass to be $\le15M_\odot$ based on nebular spectra. 
    We conclude that SN\,2024abfl is a low-luminosity SN~IIP originating from a low-mass RSG progenitor.}

    \titlerunning{SN\,2024abfl}
    \authorrunning{Xiaohan Chen, Ning-Chen Sun, and et el.}
    \maketitle

%%%%%%%%%%%%%%%%%%%%%%%%%%%%%%%%%%%%%%%%%%%%%%%%%%%%%%%%%%%%%%

\section{Introduction}
\nolinenumbers
% 添加内容用 \added{}
% 删除内容用 \deleted{}
% 替换内容用 \replaced{new}{old}

%\textcolor{red}{\textbf{physical mechanisms: Core-collapse supernovae (CCSNe), Type IIP supernovae (SNe), Red supergiant (RSG) \dots }}
Massive stars ($\gtrsim$ 8 $M_\odot$) burn hydrogen into helium during the main-sequence phase, followed by helium burning that produces a carbon-oxygen core. 
As the core continues to contract and heat up, carbon and oxygen are ignited, 
triggering a series of further nuclear reactions that ultimately form an iron core. 
The core then collapses, leading to a core-collapse supernovae (CCSNe). 
Type IIP supernovae (SNe IIP, hereafter) are a subclass of hydrogen-rich CCSNe and are characterized by a pronounced plateau in their light curves lasting $\sim$100 days \citep{1997ARA&A..35..309F,2017hsn..book..239A}
%\citep[e.g.,][]{2001ApJ...558..615H,Leonard_2002,Roy_2011,2012MNRAS.422.1122I,Huang_2016}. 
%\deleted{They are widely believed to originate from the explosions of red supergiants (RSGs)\textemdash stars with very massive hydrogen envelopes.} %\citep{1971Ap&SS..10...28G,1976ApJ...207..872C,1977ApJS...33..515F,1980ApJ...237..541A}. 
%It is such envelope that directly gives rise to the plateau observed in SNe IIP light curves. 
After core collapse, the outer envelope is ejected, heated, and ionized by the shock wave. As the ejecta expand, the recombination front moves inward in mass coordinates, causing the radius and temperature of the photosphere to remain nearly constant. This results in energy being released at an approximately constant rate, thereby producing the plateau in the light curve of SNe IIP. 
When the recombination front approaches the inner boundary of the hydrogen envelope and the entire hydrogen envelope has recombined, the luminosity drops abruptly \citep{2017hsn..book.....A}.
%\textcolor{red}{\textbf{}}
%\textcolor{red}{\textbf{the features: IIP light cureves, IIP Spectroscopy \dots}}

They are widely believed to originate from the explosions of red supergiants (RSGs)\textemdash stars with very massive hydrogen envelopes \citep{1971Ap&SS..10...28G,1976ApJ...207..872C,1977ApJS...33..515F,1980ApJ...237..541A}.
Direct imaging has confirmed that RSGs are the progenitor stars of SNe IIP \citep[e.g.][]{2003PASP..115.1289V,2016MNRAS.456L..16F,2023ApJ...955L..15N}.
However, a persistent discrepancy exists between the RSG mass range predicted by stellar evolution models and that inferred from direct detections of SNe IIP progenitors.
This inconsistency is commonly termed the "RSG problem" \citep{2009ARA&A..47...63S}. 
Specifically, modern stellar evolution models \citep{2003ApJ...591..288H,2004MNRAS.353...87E} indicate that stars with initial masses of approximately 8-25 $M_\odot$ should evolve into RSGs after completing nuclear burning and ultimately explode as CCSNe. 
However, in the direct detections of progenitors, no RSGs with initial masses > 18 $M_\odot$ have been directly linked to observed Type IIP SNe. 
Most confirmed progenitors fall within a lower mass range (concentrated around $8-16$ $M_\odot$), suggesting that higher-mass RSG progenitors appear to be "missing." 
Currently, there are primarily four explanations. 
One possible explanation is that this may result from the currently small sample size of detected SNe IIP progenitors, leading to finite sample effects \citep{2018MNRAS.474.2116D}.
Additionally, some studies point out that that the progenitor mass may be mismeasured due to uncertainties in photometric measurements such as stellar pulsation and circumstellar extinction \citep{2023ApJ...955L..15N,2024ApJ...977L..50H} or due to the blending of unresolved field stars \citep{2025ApJ...992L..24Z}.
Alternatively, the so-called failed supernova hypothesis proposes that a significant fraction of massive RSGs may not undergo successful explosions. Instead, they collapse directly into black holes, with only a small fraction of the stellar envelope being ejected \citep{1980Ap&SS..69..115N,2013ApJ...769..109L,2014ApJ...785...28K}.
A further explanation is that massive RSGs may undergo significant structural changes at certain evolutionary phases, thereby deviating from their initially predicted evolutionary paths \citep{1998A&ARv...8..145D,2012A&A...537A.146E,2012A&A...538L...8G,2025MNRAS.543.3929S}.

%\textcolor{red}{Properties of SNe IIP
%\begin{enumerate}
%    \item Inferring Type II-P Supernova Progenitor Masses from Plateau Luminosities \citep{2023ApJ...944L...2B}
%    \item Connecting the Light Curves of Type IIP Supernovae to the Properties of Their Progenitors \citep{2022ApJ...934...67B}
    %\item Observed and Physical Properties of Core-Collapse Supernovae \citep{2003ApJ...582..905H}
%\end{enumerate}
%}

Some detailed studies of the SNe IIP population have been conducted with comprehensive observations and modeling \citep{2003ApJ...582..905H,2025arXiv250620068D}.
These studies show that they cover a wide range of plateau luminosities ($-18<M_V<-13$), $^{56}$Ni mass (0.001--0.26 $M_\odot$), explosion energies (0.1--5.5 foe) and ejecta masses (5.4--24.8 $M_\odot$). 
Some SNe IIP constitute a low-luminosity subclass \citep[LLSNe~IIP;][]{2025PASP..137d4203D}. These events are characterized by narrow spectral lines (indicative of low expansion velocities), faint magnitudes ($M_{r,\mathrm{peak}} \geq -16$), low $^{56}$Ni mass (0.001--0.025 $M_\odot$) and low explosion energies (0.1--0.28 foe) \citep{10.1111/j.1365-2966.2004.07173.x,2025arXiv250620068D}. 
Studies have found that LLSNe~IIP mostly originate from low-mass RSG progenitor stars (8--12 $M_\odot$). It is worth noting, however, stars of this mass range account for $\sim$50 per cent of all massive stars, but only approximately 19 per cent of SNe~IIP explode as LLSNe~IIP \citep{2025PASP..137d4203D}. This indicates a significant diversity in the explosion of low-mass RSG supergiants, but the physical mechanisms remains unkown.
%stellar evolution and explosion mechanisms remain controversial. 
Therefore, more detailed studies of LLSNe~IIP are still needed.

SN\,2024abfl is a SN IIP discovered by Koichi Itagaki (Teppo-cho, Yamagata, Japan) on 2024 November 15.5840 UT (MJD = 60629.58) in the galaxy NGC\,2146 at an apparent magnitude of 17.5 in the clear band \citep{2024TNSTR4506....1I}, as shown in Figure~\ref{sn_images}. 
The redshift of the host galaxy is $z=0.002979$, as reported by the Transient Name Server (TNS).
Soon after discovery, \cite{2025ApJ...982L..55L} reported the direct detection of a RSG progenitor in pre-explosion images from the Hubble Space Telescope. Their analysis indicated an initial mass of $9-12$ $M_\odot$.
\cite{2026arXiv260102638G} conducted a detailed analysis of SN\,2024abfl and classified it as a LLSNe.
Therefore, SN\,2024abfl is an important target for studying LLSNe~IIP.
%\textcolor{red}{Recently, xxx reported LLSNe~IIP~IIP... Therefore, this source is an important for LLSNe~IIP~IIP...}
%Therefore, this SN serves as an important source in studying the initial mass distribution of SNe IIP progenitors. 
%, a SN IIP with a relatively long plateau phase, 

In this paper, 
we present observations of SN\,2024abfl and a detailed analysis of its light curve and spectral evolution. 
This paper is organized as follows. 
In Section~\ref{obs}, we describe our observations of SN\,2024abfl. 
Section~\ref{host} discusses the host galaxy of SN\,2024abfl. 
Section~\ref{photo} presents the photometric analysis, while Section~\ref{spec} describes the spectroscopic analysis. 
In Section~\ref{summary}, we summarize our conclusions. 

%%%%%%%%%%%%%%%%%%%%%%%%%%%%%%%%%%%%%%%%%%%%%%%%%%%%%%%%%%%%%%

%2009ib: \cite{2015MNRAS.450.3137T}
%2012aw: \cite{2013MNRAS.433.1871B}
%2013ab: \cite{2015MNRAS.450.2373B}
%2014cx: \cite{2016ApJ...832..139H}
%2015ba: \cite{2018MNRAS.479.2421D}
%2016X: \cite{2018MNRAS.475.3959H}
%2018gj: \cite{2023ApJ...954..155T}
%2020jfo: \cite{2022ApJ...930...34T}

\section{Observations}\label{obs}

\subsection{Photometry}

We conducted a multi-band photometric monitoring campaign of SN\,2024abfl with the Xinglong 60-cm telescope (XL-60), the Xinglong 35-cm telescope (XL-35), the Thai Robotic Telescope (TRT) network and the Liverpool Telescope (LT).
The observations cover the \textit{g}, \textit{r}, \textit{i}, and \textit{B}, \textit{V}, \textit{R}, \textit{I} bands.
We triggered the follow-up observations from $t=4.4$ days to $t=379.0$ days. %, initially using XL-60 and XL-35, followed by switching to TRT, and finally employing LT for the observations.
%\textcolor{red}{\textbf{
%We triggered the follow-up observations 2.1 days after discovery using XL-35 and XL-60 equipped with \textit{gri} and \textit{BVRI} filters, respectively. 
%Further observations with TRT in \textit{B}, \textit{V}, \textit{R}, \textit{I} bands were carried out starting 128.8 days after discovery.
%At 158 days after discovery, SN\,2024abfl had already entered the nebular phase. The $B$ band was no longer detectable, so we excluded this band, increased the exposure time in the $V$ band, and continued monitoring in the $VRI$ bands.
%High-cadence photometric observations were obtained up to 376.7 days after discovery.
%Our observation is from xx to xx, first use xx tel, then use xx tel}}
All the images were processed using standard reduction procedures, including bias subtraction and flat-field correction. 
Point-spread-function (PSF) photometry was then performed with the Automated Photometry of Transients (AUTOPHOT) package \citep{2022A&A...667A..62B}. 
Since SN\,2024abfl is located in the outskirts of its host galaxy, where the host contribution is negligible, 
we perform direct photometry without image subtraction.
%we did not perform image subtraction. 
%Instead, photometry was carried out directly after the image pre-processing.

%\begin{figure}[ht!]
%\centering
%\includegraphics[width=\hsize]{figures/sn_image_zoom_in.eps}
%    \caption{SN 2024abfl and its host galaxy NGC\,2146. 
%    Shown is an RGB composite assembled from \replacedd{Xinglong 35-cm telescope}{XL-35} \textit{g}-, \textit{r}-, and \textit{i}-band exposures at t = \replacedd{4.4}{2.1} days after the discovery.}
%        \label{sn_images}
%\end{figure}

\begin{figure}[ht!]
    \centering
    \includegraphics[width=\hsize]{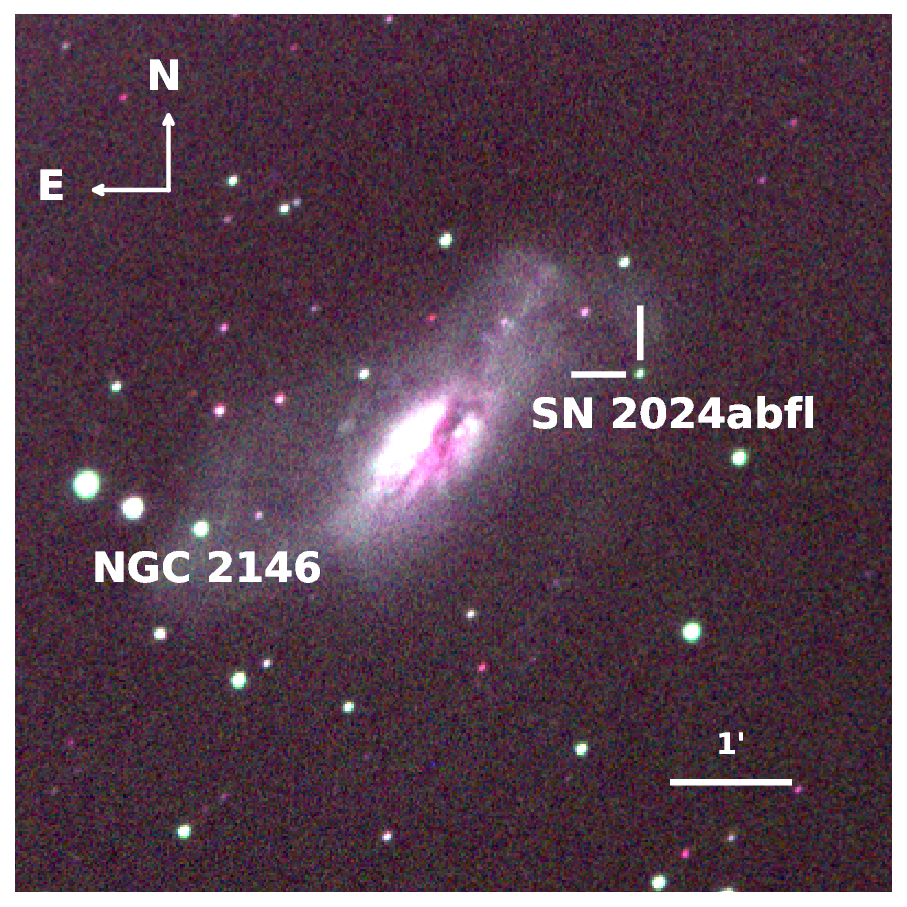}
        \caption{SN 2024abfl and its host galaxy NGC\,2146. 
        Shown is an RGB composite assembled from Xinglong 35-cm telescope \textit{g}-, \textit{r}-, and \textit{i}-band exposures at t = 4.4 days after the explosion.}
            \label{sn_images}
    \end{figure}

\begin{table}[ht!]
\caption {Ultraviolet-optical photometric results of SN\,2024abfl. We only show the first 15 rows; the full table is available online. 
Magnitudes are reported in the default photometric systems: Johnson-Cousins \textit{BVRI} and Swift/UVOT bands are calibrated in the Vega system, whereas SDSS \textit{gri} and ATLAS \textit{c} and \textit{o} bands are calibrated in the AB system. The phase represents days relative to the date of explosion, $t_0=60627.27$ MJD.}
\label{phot_log} 
\centering
\renewcommand{\arraystretch}{1.4}
\addtolength{\tabcolsep}{-3pt}
\begin{tabular}{c c c c c c}
\hline\hline             
MJD & Phase & Filter & Magnitude & Uncertainty & Instrument \\         % table heading
\hline                      % inserts single horizontal line
    60628.2 & 1.0 & \textit{g} & 17.26 & 0.22 & ZTF\\    % inserting body of the table
    60628.3 & 1.0 & \textit{r} & 17.44 & 0.03 & ZTF\\
    %60629.5 & 0.0 & \textit{o} & 17.22 & 0.07 & ATLAS\\
    %60629.6 & 0.0 & \textit{o} & 17.14 & 0.06 & ATLAS\\
    60629.6 & 2.3 & \textit{o} & 17.17 & 0.05 & ATLAS\\
    60631.3 & 4.0 & \textit{r} & 17.20 & 0.10 & ZTF\\
    60631.5 & 4.2 & \textit{o} & 16.98 & 0.03 & ATLAS\\
    %60631.5 & 1.9 & \textit{o} & 19.09 & 0.09 & ATLAS\\
    %60631.5 & 1.9 & \textit{o} & 16.86 & 0.08 & ATLAS\\
    %60631.5 & 1.9 & \textit{o} & 17.02 & 0.08 & ATLAS\\
    %60631.5 & 1.9 & \textit{o} & 16.75 & 0.15 & ATLAS\\
    %60631.5 & 1.9 & \textit{o} & 17.05 & 0.08 & ATLAS\\
    %60631.5 & 1.9 & \textit{o} & 17.00 & 0.09 & ATLAS\\
    %60631.7 & 2.1 & \textit{o} & 16.91 & 0.15 & ATLAS\\
    60631.7 & 4.4 & \textit{i} & 17.03 & 0.06 & XL-35\\
    60631.7 & 4.4 & \textit{g} & 16.99 & 0.04 & XL-35\\
    60633.0 & 5.7 & \textit{B} & 17.37 & 0.30 & XL-60\\
    60633.7 & 6.4 & \textit{i} & 16.82 & 0.08 & XL-35\\
    60633.7 & 6.4 & \textit{g} & 17.11 & 0.06 & XL-35\\
    60634.0 & 6.7 & \textit{B} & 16.94 & 0.23 & XL-60\\
    60634.0 & 6.7 & \textit{V} & 16.69 & 0.19 & XL-60\\
    60634.0 & 6.7 & \textit{R} & 16.6 & 0.12 & XL-60\\
    60634.0 & 6.7 & \textit{I} & 16.41 & 0.14 & XL-60\\
    60634.7 & 7.4 & \textit{i} & 17.23 & 0.07 & XL-35\\
\hline
\end{tabular}
\end{table}

The galaxy NGC\,2146 also hosted SN\,2018zd, a previous event proposed to have an electron-capture origin \citep{2021NatAs...5..903H}. 
Swift's long-term monitoring campaign of SN\,2018zd serendipitously covered the first few days after the explosion of SN\,2024abfl. 
%The Swift conducted long-term monitoring observation of SN\,2018zd over several years with several epochs of the observations carry out after the explosion of SN\,2024abfl.
%The UltraViolet/Optical Telescope (UVOT) onboard the Neil Gehrels Swift Observatory monitored SN\,2018zd over an extended period, 
%\textcolor{red}{\textbf{Maybe for observation of X-ray, swift observe; long-term observation over several years}}
%\textcolor{red}{\textbf{with several epochs of the observations carry out after the explosion of SN\,2024abfl.
%cover ... the explosion
% continuing through the epoch of the SN\,2024abfl explosion. }}
We retrieved the corresponding UVOT data obtained with the \textit{UVW2}, \textit{UVM2}, and \textit{UVW1} filters \footnote{\url{https://www.swift.ac.uk/archive/selectseq.php?tid=00010592&source=obs&name=AT2018zd&reproc=1&referer=portal}}.
Aperture photometry was performed with HEASoft/uvotsource, adopting a 3" extraction radius. 
%Since the data included pre-explosion images of SN\,2024abfl, 
Flux subtraction of the host background was applied based on the pre-explosion images.
%\textcolor{red}{\textbf{website, arXiv, comform subtraction, sub flux}}

We also obtained ATLAS and ZTF data for SN\,2024abfl using their forced-photometry services \citep{2018PASP..130f4505T, Masci_2019}. 
All photometric results are listed in Table~\ref{phot_log}.

\subsection{Spectroscopy} 

\begin{table*}[ht!]
\caption {Spectroscopic observation of SN\,2024abfl. The phase represents days relative to the date of explosion, $t_0=60627.27$ MJD.}
\label{spec_log} 
\centering
\renewcommand{\arraystretch}{1.5}
\begin{tabular}{ccccc}
\hline\hline             
MJD & Phase (d) & Filter and Grism & Spectral range (\AA) & Telescope/Instrument\\
\hline
    60629.9 & 2.6 & 676R$+$420 gr/mm & 3500-8950 & HCT/HFOSC\\
    60630.5 & 3.2 & B480$+$G5309 & 3750-7500 & Gemini-N/GMOS\\
    60630.9 & 3.6 & 385LP$+$G4 & 3700-8800 & XL-216/BFOSC\\
    60631.9 & 4.6 & 385LP$+$G4 & 3700-8800 & XL-216/BFOSC\\
    60658.6 & 31.4 & 385LP$+$G4 & 3700-8800 & XL-216/BFOSC\\
    60661.7 & 34.5 & 385LP$+$G4 & 3700-8800 & XL-216/BFOSC\\
    60666.9 & 39.6 & 385LP$+$G4 & 3700-8800 & XL-216/BFOSC\\
    60667.9 & 40.6 & 385LP$+$G4 & 3700-8800 & XL-216/BFOSC\\
    60675.6 & 48.3 & 385LP$+$G4 & 3700-8800 & XL-216/BFOSC\\
    60678.8 & 51.5 & 385LP$+$G4 & 3700-8800 & XL-216/BFOSC\\
    60680.7 & 53.4 & 385LP$+$G4 & 3700-8800 & XL-216/BFOSC\\
    60690.7 & 63.5 & 385LP$+$G4 & 3700-8800 & XL-216/BFOSC\\
    60714.6 & 87.3 & 385LP$+$G4 & 3700-8800 & XL-216/BFOSC\\
    60719.5 & 92.2 & 385LP$+$G4 & 3700-8800 & XL-216/BFOSC\\
    60722.5 & 95.3 & 385LP$+$G4 & 3700-8800 & XL-216/BFOSC\\
    60746.6 & 119.3 & 385LP$+$G4 & 3700-8800 & XL-216/BFOSC\\
    60767.5 & 140.3 & 385LP$+$G4 & 3700-8800 & XL-216/BFOSC\\
    %60784.9 & 155.3 &  & Grism \#4 & 3200-9600 & NOT/ALFOSC\\
    60816.9 & 189.6 & Grism \#4 & 3200-9600 & NOT/ALFOSC\\
    60825.9 & 198.6 & Grism \#4 & 3200-9600 & NOT/ALFOSC\\
\hline
\end{tabular}
\end{table*}

Spectroscopic monitoring of SN\,2024abfl was conducted from t = 3.6 days to t = 198.6 days with two facilities: the Xinglong 2.16-m telescope (XL-216) and 2.56-m Nordic Optical Telescope (NOT).
The spectral reduction was performed with standard IRAF routines, involving bias subtraction, flat-field correction, cosmic-ray removal, flux calibration and wavelength calibration.

Additionally, two publicly available spectra were retrieved from TNS: one obtained with the Frederick C. Gillett Gemini North Telescope (Gemini-N) at t = 3.2 days, and the other with the 2.01-m Himalayan Chandra Telescope (HCT) at t = 2.6 days \citep{2024TNSCR4515....1D,2024TNSCR4535....1A}.
A log of all spectroscopic observations is listed in Table~\ref{spec_log}.

%%%%%%%%%%%%%%%%%%%%%%%%%%%%%%%%%%%%%%%%%%%%%%%%%%%%%%%%%%%%%%

\section{Host Galaxy}\label{host}

SN\,2024abfl is located in the outskirts of NGC\,2146, a nearby starburst galaxy that has undergone a recent merger event \citep{2012AAS...21943803A}. 
In this section, we discuss the distance, metallicity, and extinction of the host galaxy, which are essential for the analysis of SN\,2024abfl.

\subsection{Distance}

%\textcolor{red}{Superbol: Distance}

%\textcolor{red}{Fred: How do this range of distances impact on the derived mass of the progenitor ? (email) WuJunJie: Luo.等年龄线}

%\textcolor{red}{Fred: Just a comment - From the conclusions of Luo et al. (2025)  (section 5): Currently, we understand that the biggest part of the error came from the large scatter in the host galaxy distance measurement; we hope deep HST or JWST observation in the future will help determine the distance of the host galaxy via Cepheid variables or TRGB measurements.}

The NASA/IPAC Extragalactic Database (NED) lists 16 distance measurements for NGC\,2146, of which 13 are based on the Tully-Fisher method \citep{1983A&A...118....4B}, one on the Sosies method \citep{2002A&A...393...57T}, and the remaining two on the tip of the red giant branch (TRGB) method and globular-cluster radius (GC radius) technique \citep{2012MNRAS.426.1185A}, respectively.
Distances derived with the Tully-Fisher method span 10.3-39.7 Mpc, 
the Sosies approach gives 27.7 Mpc, and both the TRGB and GC radius techniques yield 18 Mpc. 
Because NGC\,2146 underwent a strong tidal interaction with another galaxy within the last Gyr, the scatter of the Tully-Fisher relation becomes significantly larger for such an interacting system \citep{callis2021lowgosn2018zd}. 
For NGC\,2146, we adopt the $15.6_{-3.0}^{+6.1}$ Mpc distance provided by the combined probability distribution of kinematic, Tully-Fisher and GC-radius measurements \citep{callis2021lowgosn2018zd}.
%\citep{2012MNRAS.426.1185A}

\subsection{Metallicity}

\cite{2019A&A...623A...5D} used literature and VLT/MUSE emission line fluxes to derive gas-phase metallicities (oxygen abundances) for more than 10,000 individual regions, 
and they also determined characteristic metallicities for each galaxy. 
For NGC\,2146, they reported a global gas-phase metallicity of $12+\log(\text{O/H})=8.78\pm0.16$ \footnote{\url{http://dustpedia.astro.noa.gr/AncillaryData}}, equal to 1.2 Z$_\odot$.
SN\,2024abfl is situated in the outskirts of NGC\,2146 (as shown in Figure \ref{sn_images}), where the metallicity is expected to be lower.

%\textcolor{red}{Explosion epoch: the SuperNova IDentification package (SIND,\cite{2007ApJ...666.1024B})}
%\textcolor{red}{Distance and Explosion epoch: the expanding photosphere method (EPM, application:\cite{2018MNRAS.479.2421D})}

%\textcolor{red}{ZY: estimate the metallicity at the SN position. Please see Table 4 in doi:10.1093/mnras/stv335 for reference.}

\subsection{Extinction}

To obtain a more accurate understanding of SN\,2024abfl, it is essential to estimate the total line-of-sight extinction toward it.
The Galactic extinction toward NGC\,2146 is $A_V^\text{MW}$ = 0.26 mag according to \cite{2011ApJ...737..103S}. 
For host extinction, there are two commonly used methods for a rough estimation: (i) the equivalent width (EW) of the Na \textsc{i} D absorption line, (ii) the $(V-I)$ colours.%, (iii) a comparison of the early spectral shape with a blackbody function.}

\begin{figure}[ht!]
    \centering
    \includegraphics[width=\hsize]{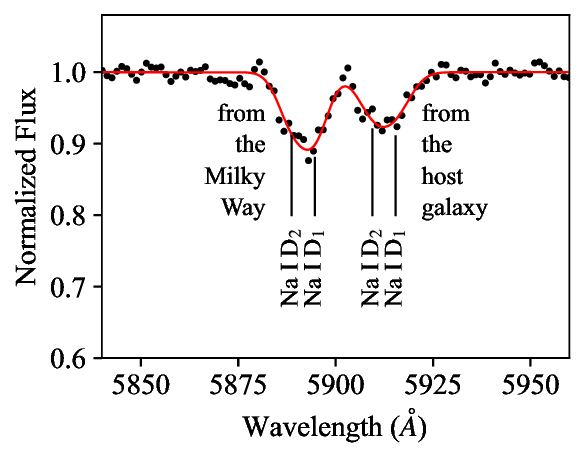}
        \caption{Normalized Gemini spectrum of SN 2024abfl at t = 3.2 day, showing prominent Na I D absorption features. The observed data (black points) are fitted with Gaussian profiles (red curves). The left absorption component arises from the Milky Way, while the right one is associated with the host galaxy. }
            \label{NaID}
    \end{figure}

First, we used the Gemini-N spectrum obtained at t = 0.9 days to estimate the host-galaxy extinction by measuring the EW of the Na \textsc{i} D absorption.
As shown in Figure~\ref{NaID}, two absorption components are present: one near the rest wavelengths of the doublet and another at a higher redshift of $z=0.0033$. 
We attribute the former component to the Milky Way and the latter to NGC\,2146. 
Since the Na \textsc{i} D doublet is superimposed with a broad He \textsc{i} emission, the baseline was modeled using a fourth-order polynomial and each absorption feature was fitted with a Gaussian profile. The total EW of the host-galaxy Na \textsc{i} D doublet is EW$^\text{Host}$ = 0.5 \AA + 0.52 \AA = 1.02 \AA.
The empirical relation between the EW of Na \textsc{i} D and the colour excess $E(B-V)$ derived by \cite{2012MNRAS.426.1465P} is employed to calculate the the colour excess $E(B-V)$.
For simplicity, we assume the standard extinction law with $R_V$ = 3.1 \citep{1989ApJ...345..245C}.
This yields a host extinction of $A_V^\text{Host} = 0.68\pm0.12$ mag. 

We also use the `colour method' to estimate the host extinction \citep{2010ApJ...715..833O}. This method assumes that all SNe~IIP have the same photospheric temperature at the end of the plateau phase ($t_\mathrm{PT}-30$ days), and that color differences are primarily due to host galaxy dust extinction. The calculation formula is as follows:
\begin{align}
    &A_V(V-I) = 2.518 \left[ (V - I) - 0.656 \right], \\
    &\sigma(A_V) = 2.518 \sqrt{ \sigma_{(V-I)}^2 + 0.0053^2 + 0.0059^2 }.
\end{align}
We estimated the $(V - I)$ colours (corrected for Galactic extinction) at the end of the plateau phase ($\sim100$ days) and obtained a result of $A_V^\mathrm{Host}=0.77 \pm 0.51$, which is consistent with the value obtained using the first method within uncertainties.

%\addedd{Third, \textsc{snid} code \citep{2007ApJ...666.1024B}}

Ultimately, we adopt the $A_V^\text{Host} = 0.68\pm0.12$ mag throughout the paper.
We note that, however, the result of \cite{2026arXiv260102638G} ($E(B-V)_\mathrm{host}=0.01\pm0.08$ mag) is lower than ours. 

\section{Photometric Analysis}\label{photo}

\subsection{Main light-curve features}

\begin{figure*}[ht!]
    \centering
        \resizebox{17.5cm}{10.5cm} % 5:3
    {\includegraphics {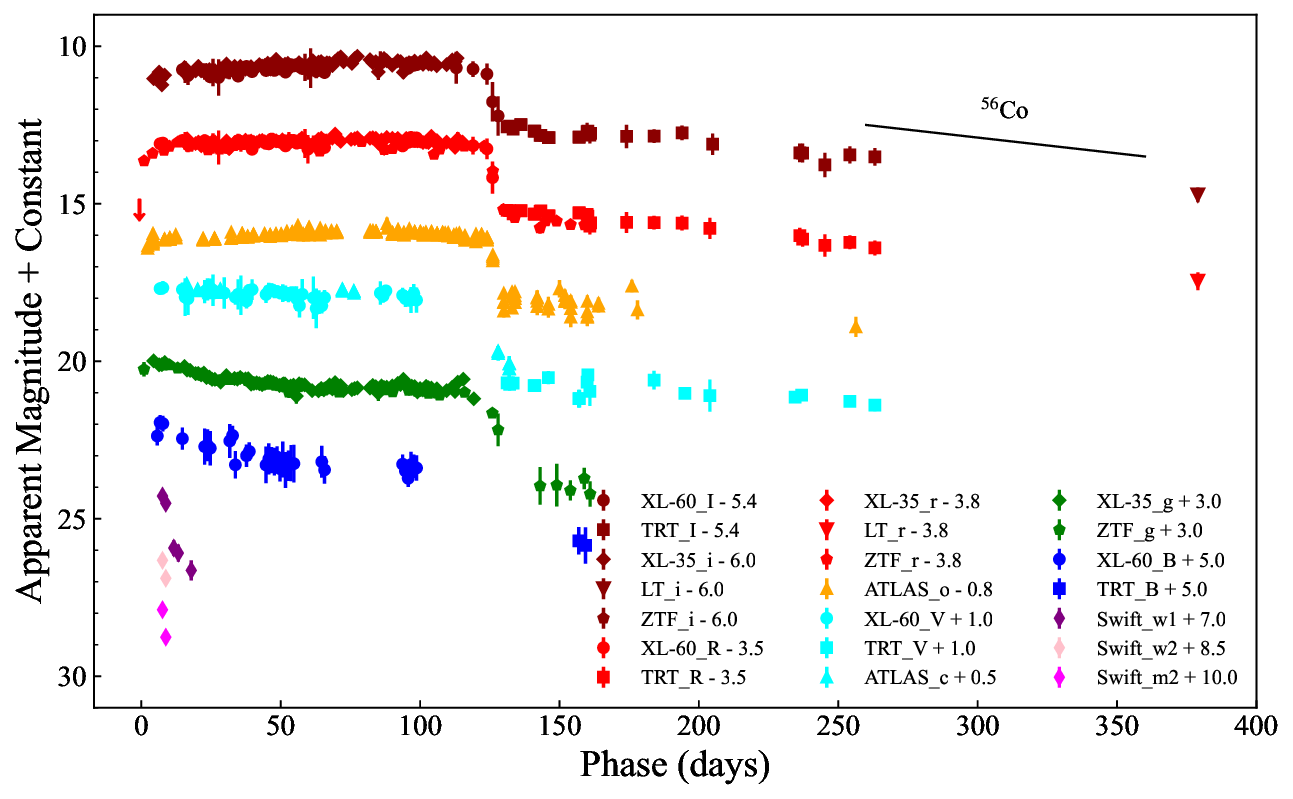}}
        \caption{The multi-band light curves of SN\,2024abfl. The phase represents days relative to the date of explosion, $t_0=60627.27$ MJD.}%Phases are marked relative to the first clear-band detection (MJD 60629.58).}
        \label{multiband_lc}
\end{figure*}

The multi-band light curve of SN\,2024abfl is shown in Figure~\ref{multiband_lc}. 
The last non-detection of SN\,2024abfl in the ZTF\_\textit{r} band was on $t=-0.96$ day, followed by a clear detection on $t=0.97$ day in the ZTF\_\textit{g} band.
We assume the explosion time of SN\,2024abfl to be the median of these two epochs, i.e. $t=0.0$ $\pm$ 0.97 day. 
Shortly afterward, Koichi Itagaki reported it at MJD = 60629.58 ($t=2.31$ days). 
Unless otherwise noted, all epochs in this paper are given relative to the explosion date.
%\textcolor{red}{(Why are you using the discovery date as a reference? With a last non detection followed by a detection less than 2 days later, I'd say you should use the averaged epoch as the estimate for the explosion date, which is more commonly used for SNe II-P and is more physical.)}
In the first few days, the \textit{UV}-band light curves decline sharply, which indicates a rapid cooling of the SN photosphere during the early phase.
At the same time, the \textit{g-} and \textit{B-}band light curves show a slight decline,
whereas the \textit{o-}, \textit{r-}, \textit{R-}, \textit{i-}, and \textit{I-}band light curves exhibit a slight rise. 

\begin{figure}[ht!]
\centering
\includegraphics[width=\hsize]{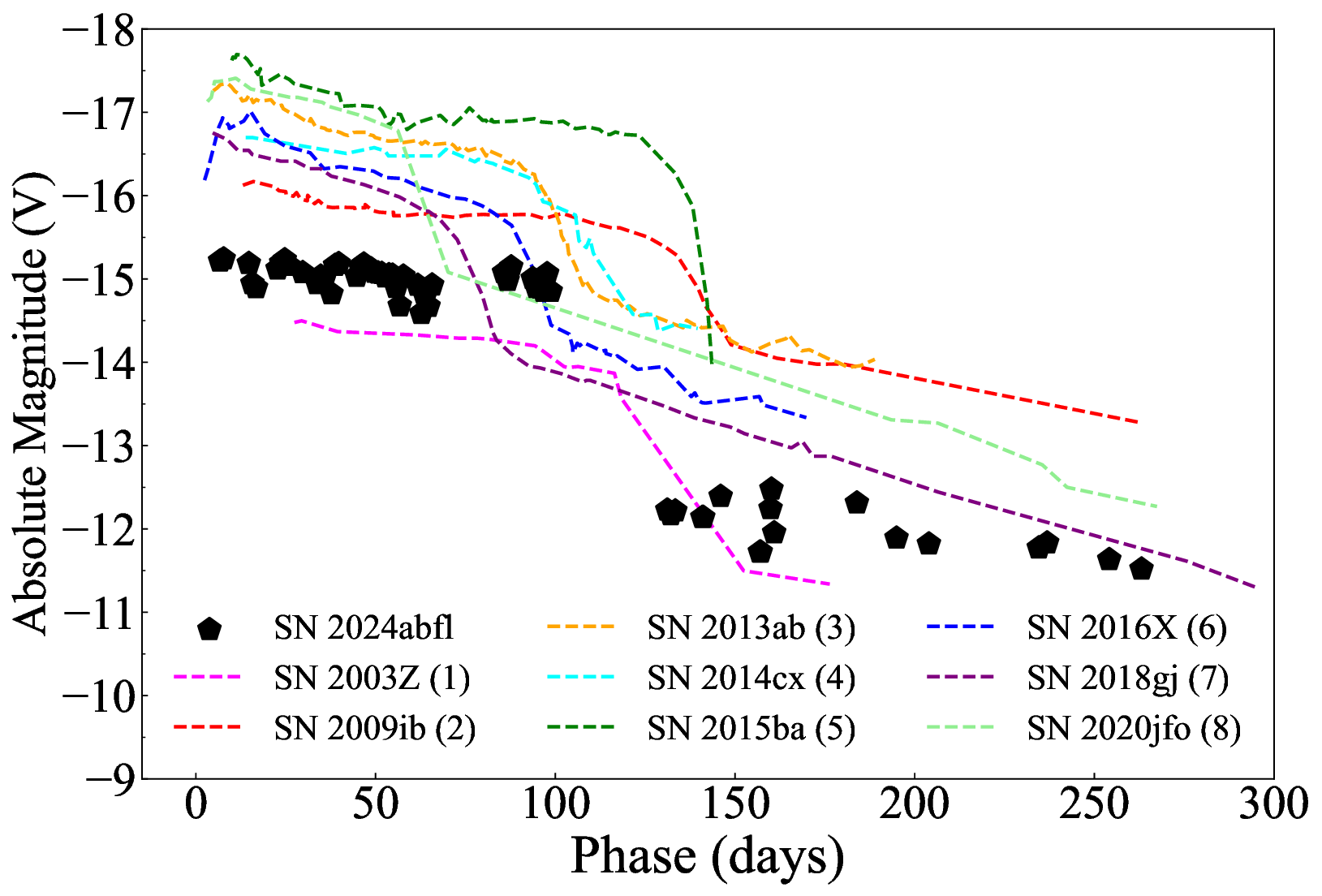}
    \caption{The \textit{V}-band light curve of SN\,2024abfl compared with other SNe IIP. 
    (1) \cite{2014MNRAS.439.2873S},
    (2) \cite{2015MNRAS.450.3137T}, 
    %(2) \cite{2013MNRAS.433.1871B}, 
    (3) \cite{2015MNRAS.450.2373B}, 
    (4) \cite{2016ApJ...832..139H}, 
    (5) \cite{2018MNRAS.479.2421D},
    (6) \cite{2018MNRAS.475.3959H},
    (7) \cite{2023ApJ...954..155T},
    (8) \cite{2022ApJ...930...34T}.}
        \label{R_band_compare}
\end{figure}

The plateau has a typical magnitudes of $M_V$ $\sim-15$ mag. 
During the plateau, the $r$-band magnitude of SN\,2024abfl declines by less than 0.5 mag, and its $B$-band decline rate is 1.28 mag per 100 days, consistent with the characteristics of typical Type IIP supernovae \citep{2011MNRAS.412.1441L,1994A&A...282..731P}.
We estimated the plateau length by fitting the analytic function provided in \cite{2016MNRAS.459.3939V}, as shown below:
\begin{equation}
    y(t)=\frac{-a_0}{1+e^{(t-t_\text{PT})/w_0}}+(p_0\times t)+m_0
\end{equation}
where $t$ is the time in days relative to the explosion; 
$t_{\mathrm{PT}}$ is the time in days from explosion to the transition point between the end of the plateau phase and start of the radioactive tail; 
$a_0$ represents the depth of the drop in magnitudes from the plateau to the radioactive tail; 
$w_0$ describes the slope of this drop; 
$p_0$ characterizes the slope on either side of the drop and $m_0$ is a constant. 
Setting the explosion time as $t = 0$, we fitted this function to the $V$-band light curve of SN\,2024abfl, 
obtaining a best-fitting value of $t_\mathrm{PT}=126.54 \pm 0.65$ days. % and $a_0$ $\sim$ 2.6 mag. 
Figure~\ref{R_band_compare} shows the \textit{V}-band light curve of SN\,2024abfl compared with other SNe IIP, showing that SN\,2024abfl exhibits a relatively long plateau phase and low luminosity compared with most SNe IIP \citep{10.1111/j.1365-2966.2004.07173.x}. 
%\added{The peak magnitudes of \textit{V}-band is greater than -15.5 mag, which is  }

The plateau was followed by a sharp decline over the subsequent $\sim$6 days, 
corresponding to a $\sim$2-magnitude drop in the \textit{r} band from the plateau to the radioactive tail.
All light curves decline slowly in the radioactive tail with a rate of $\sim$0.9 mag per 100 days.

%SN\,2024abfl was first reported approximately 2 days after explosion, 
%at a stage when it was descending from the peak and entering the plateau phase. 
%The light curve of SN\,2024abfl shows a prominent plateau after maximum, during which the brightness declines slowly. 
%The plateau length of SN\,2024abfl (about 120.8 days in the V band) is similar to that of SN\,2015ba (about 123 days in the V band; \cite{2018MNRAS.479.2421D}), 
%but the magnitude drop from the plateau to the nebular phase is somewhat smaller for SN\,2024abfl (about 2.7 mag). 

%\textcolor{red}{爆发时刻；
%按照时间顺序描述：last-nondetection,first-deteciton; 
%first peak(按照波段，in the first few days, g-band..., B-band....,uv-band...表明peak，温度;); 
%第二段：plateau持续时间; sharply decline; radioactive tail, rate; 
%第三段：进一步，拟合和比较
%爆发时刻and first few days、
%平台and decline and compare with other SNe and tail(斜率)}

\subsection{Bolometric Light Curve}

\begin{figure}[ht!]
\centering
\includegraphics[width=\hsize]{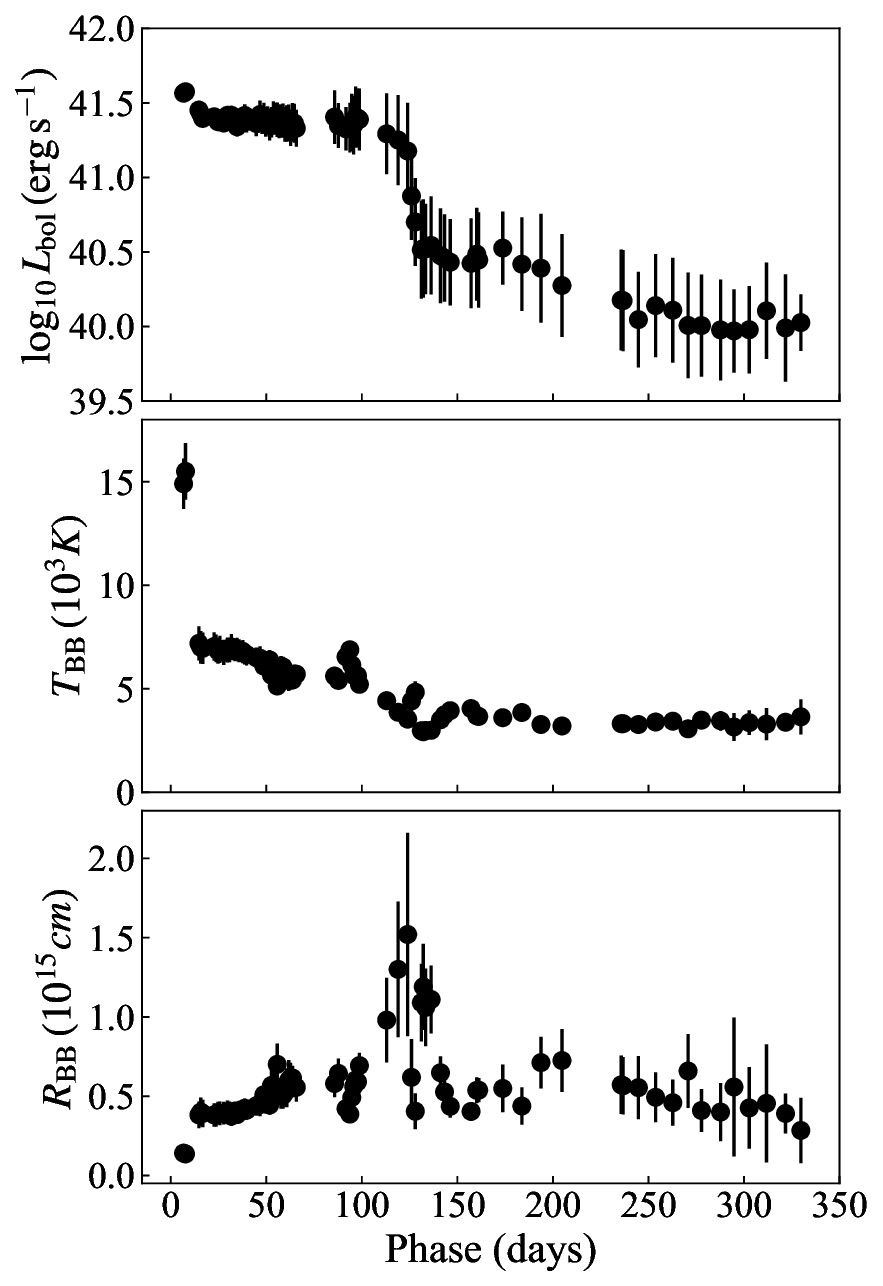}
    \caption{\textit{Top panel}: Logarithmic bolometric luminosity ($L_{\mathrm{bol}}$) from observed fluxes.
    %ed logarithmic bolometric luminosity ($L_{\mathrm{bol}}$) (red diamonds).
    \textit{Middle panel}: Evolution of the blackbody temperature ($T_{\mathrm{BB}}$).
    \textit{Bottom panel}: Evolution of the blackbody radius ($R_{\mathrm{BB}}$).
    All parameters are shown as a function of days since explosion (MJD 60627.27).}
        \label{bolometric_lc}
\end{figure}

We analyzed the evolution of the bolometric luminosity ($L_\mathrm{bol}$), blackbody temperature ($T_\mathrm{BB}$), and radius ($R_\mathrm{BB}$) of SN\,2024abfl using the SuperBol package \citep{2018RNAAS...2..230N}. Our analysis incorporated Swift UVOT ($uvw2$, $uvm2$, $uvw1$) and ground-based optical ($B$$V$$R$$I$$g$$r$$i$) photometry, with light curves in each band interpolated to the temporal sampling of the reference band and extrapolated where necessary assuming constant colors. 
As shown in Figure~\ref{bolometric_lc}, 
the luminosity exhibits an initial peak of $\sim3.8\times10^{41}$ erg s$^{-1}$, accompanied by a high temperature of $\sim1.5\times10^4$ K and a small radius of $\sim1.4\times10^{14}$ cm. 
As the shock-heated SN ejecta expands and cools, over the next $\sim10$ days, the luminosity then decreases rapidly to $\sim2.5\times10^{41}$ erg s$^{-1}$, the temperature also exhibits a rapid decline to $\sim7000$ K, and the radius increases to $\sim3.9\times10^{14}$ cm.
%following core collapse, a shock wave propagates into the hydrogen envelope, ejecting, heating, and ionizing the envelope material. 

During the plateau phase, the sustained luminosity is primarily powered by radiative diffusion of the thermal energy released during the recombination of ionized hydrogen in the shock-heated envelope. 
Over this period, the temperature slightly decreases to $\sim5000$ K, 
%$\sim4390$ K, 
while the radius slightly expands to $\sim5\times10^{14}$ cm. 
%$\sim1.60\times10^{15}$ cm. 

%Notably, as the plateau phase ends and the light curve enters the radioactive tail, the uncertainty in the radius becomes significantly large. 
%Around day 121, the light curve exhibits a pronounced and abrupt decline. 
After the plateau, 
the luminosity drops sharply to $\sim3.3\times10^{40}$ erg s$^{-1}$ over a timescale of approximately 7 days, which marks the end of hydrogen recombination within the ejecta. 
Following this steep decline, SN\,2024abfl enters the radioactive-decay tail, where the light curve is primarily powered by the energy deposited from the radioactive decay of $^{56}$Co to $^{56}$Fe. 
In this phase, the temperature remains roughly stable at $\sim3000$ K, while the radius gradually decreases to $\sim3\times10^{14}$ cm. 
%The luminosity initially declines rapidly from a peak of 
%$\sim6.30\times10^{41}$ erg s$^{-1}$ to 
%$\sim4.21\times10^{41}$ erg s$^{-1}$, 
%after which it enters a remarkably slow and extended decline lasting approximately 120 days. 
%By $\sim$121 days, the luminosity has decreased to 
%$\sim3.37\times10^{41}$ erg s$^{-1}$, 
%corresponding to a modest reduction of only 
%$\sim0.84\times10^{41}$ erg s$^{-1}$. 
%This prolonged plateau phase is a defining feature of Type IIP supernova light curves. 

% within the first $\sim$12 days post-discovery. 
%This is followed by a more gradual 
%cooling to $\sim$4390 K over the next $\sim$111 days, 
%and finally stabilizes after a very slow decline to $\sim$3120 K.

%The radius increases from approximately $9.39\times10^{13}$ cm to $1.60\times10^{15}$ cm by $\sim$129 days, 
%then decreases to $\sim$$2.40\times10^{14}$ cm in the next 197 days.

%\textcolor{red}{non-detection to peak: the sn has a initial peal of ..., decline ... days.
%按照阶段将，post shock, 半径小，温度高，亮；平台期不太变化slit decline, H recombination; 
%radius误差比较大；tail阶段...温度稳定，另外缓慢下降；结合物理过程}

%\textcolor{red}{\textbf{------------------------2025.12.23----------------------}}

\subsection{$^{56}$Ni mass from the radioactive tail}

%\begin{equation}
%    \log M(^{56}Ni)=-(3.5024 \pm 0.0960) \times S- 1.0167 \pm 0.0034.
%\end{equation}
%$S=0.0631 \pm 0.0044$ mag d$^{-1}$,  $\mathrm{M}(^{56}Ni)=0.058\pm0.0004$
$^{56}$Ni mass can be independently estimated from the tail $L_\mathrm{bol}$ 
by the following equation \citep{2003ApJ...582..905H}, 
which back-traces the observed $L_\mathrm{bol}$ through the  
$^{56}$Ni$\to$$^{56}$Co$\to$$^{56}$Fe decay chain to the amount of nickel originally synthesized:
\begin{equation}
    M_{Ni}=7.866\times10^{-44}\times L_\mathrm{t}\exp{\frac{(t-t_0)/(1+z)-6.1}{111.26}}M_\odot
\end{equation}
where $z$ is the redshift, 6.1 is the half-life of $^{56}$Ni in days, 111.26 is the e-folding time of the $^{56}$Co decay in days, and $t_0$ is the explosion time. 
The tail luminosity of SN\,2024abfl, $L_\mathrm{t}=1.5\times10^{40}$ erg s$^{-1}$, measured at $\sim235.73$ days, implies a $^{56}$Ni mass of $\sim0.009$ $M_\odot$. 
%bolometric correction (BC) factor of $0.26 \pm 0.06$ \citep{2001ApJ...558..615H}
%\begin{equation}
%    \log(L_t)=-0.4[V_t- A_v(tot) + \textsc{BC}] + 2\log D- 3.256
%\end{equation}

\subsection{Analytical modelling}

%\textcolor{red}{New Superbol, new model}

We also estimate the $^{56}$Ni mass, the initial radius, the ejecta mass and the explosion energy of SN\,2024abfl using the Markov Chain Monte Carlo (MCMC) version of the semi-analytic light curve code of \cite{2014A&A...571A..77N} and \cite{2020MNRAS.496.3725J}. 
%The model assumes a spherically symmetric SN ejecta with the diffusion approximation, and a simplified treatment of recombination, while neglecting circumstellar medium (CSM) interaction and binary effects, making it suitable for rapid estimation of the basic physical parameters. 
The model assumes homologously expanding, spherically symmetric ejecta of uniform density. 
% core and an outer envelope with an exponential density profile (the shape parameters may be set to zero, reducing the ejecta to a single uniform sphere). 
Radiation transport is treated under the diffusion approximation, and the opacity is assumed to be piecewise constant, with $\kappa = 0.2~\mathrm{cm}^2\mathrm{g}^{-1}$ in the ionized region and $\kappa = 0~\mathrm{cm}^2\mathrm{g}^{-1}$ in the recombined, neutral region, separated by a recombination front propagating inward \citep{2016A&A...589A..53N,2020MNRAS.496.3725J}. 
The luminosity is powered solely by the $^{56}\mathrm{Ni} \rightarrow {}^{56}\mathrm{Co} \rightarrow {}^{56}\mathrm{Fe}$ radioactive decay chain, while contributions from circumstellar interaction or a binary companion are neglected. The $\gamma$-ray optical depth is parameterized as $\tau_\gamma = A_g / t^2$, where the parameter $A_g$ controls the efficiency of energy deposition during the radioactive tail and, together with the nickel mass, determines the late-time luminosity. 

\begin{figure}[ht!]
\centering
\includegraphics[width=\hsize]{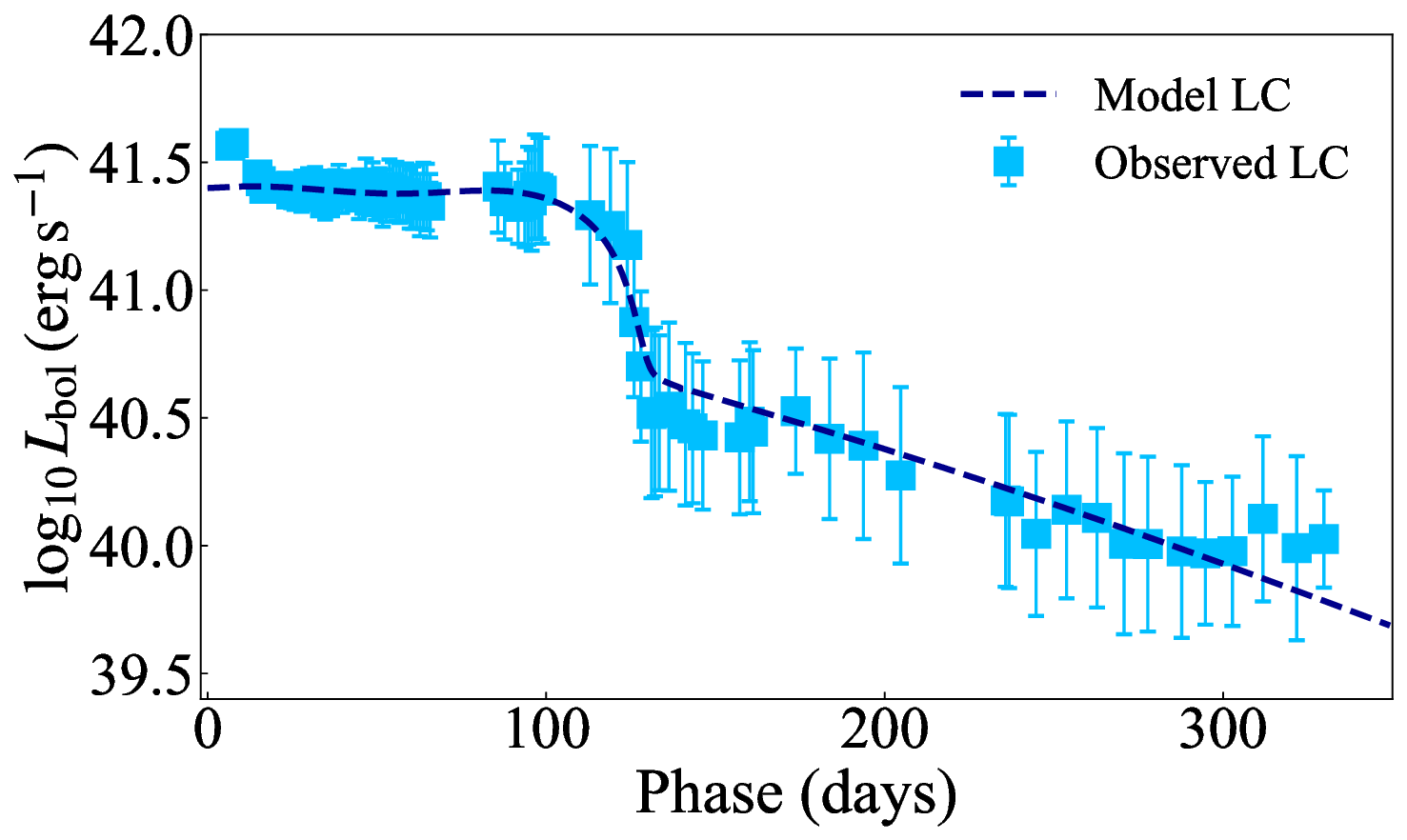}
    \caption{The best-fitting analytical model light curve of SN\,2024abfl using \cite{2014A&A...571A..77N} and \cite{2020MNRAS.496.3725J}. The phase represents days relative to the date of explosion, $t_0=60627.27$ MJD.}
    %\textcolor{red}{The derived model parameters from the fitting are $E = 1.0025$ foe, $R=6.5\times10^{13}$ cm,and $M_\text{ej} =12.0$ M$_\odot$.}}
        \label{analytical_mcmc_modelling}
\end{figure}

We fit this model to the bolometric light curve of SN\,2024abfl, 
exploring a parameter space defined by 
$R_0 \in (1$--$10)\times10^{13}$ cm (the typical radius of RSGs), 
$M_\mathrm{ej} \in 4$--$20$ $M_\odot$, 
$E_\mathrm{kin} \in$ 0.001--1 foe (1 foe $=10^{51}$ erg), 
$E_\mathrm{th} \in 0.001$--$1$ foe, 
and $M_\mathrm{Ni} \in 0.0005$--$0.25$ $M_\odot$ (see Table~\ref{analytical_modelling_parm} for the explanations of these parameters). 
The resulting best-fitting parameters are listed in Table~\ref{analytical_modelling_parm} and
the best-fitting light curve is illustrated in Figure~\ref{analytical_mcmc_modelling}. 
%The best-fitting explosion energy, progenitor radius and ejecta mass are 0.759 feo, $4.924\times10^{13}$ cm and 9.197 M$_\odot$, respectively. 
The estimated $^{56}$Ni mass is $\sim0.009$ $M_\odot$, consistent with the value of $\sim0.009$ $M_\odot$ derived from the tail bolometric luminosity. 
The $^{56}$Ni mass, explosion energy, and ejecta mass are much lower than the typical range for normal SNe IIP ($^{56}$Ni mass: 0.001--0.26 $M_\odot$; explosion energies: 0.1--5.5 foe; ejecta mass: 5.4--24.8 $M_\odot$) but consistent with LLSNe~IIP ($^{56}$Ni mass: 0.001--0.025 $M_\odot$; explosion energies: 0.1--0.28 foe) \citep{10.1111/j.1365-2966.2004.07173.x,2025PASP..137d4203D,2025arXiv250620068D}.
%These results fall within the range of LLSNe~IIP ($^{56}$Ni mass: 0.001--0.025 $M_\odot$; explosion energies: 0.1--0.28 foe; progenitor mass: 8--12 $M_\odot$), although the explosion energy is slightly above the typical range for LLSNe~IIP \citep{10.1111/j.1365-2966.2004.07173.x,2025PASP..137d4203D,2025arXiv250620068D}.
It is noteworthy that our model does not fit the first peak, which as \cite{2026arXiv260102638G} suggested is caused by early-time circumstellar medium (CSM) interaction in SN\,2024abfl.

%\textcolor{red}{Our results's features: lower than IIP, consistent with typical value.}

%\textcolor{red}{compare Ni}
%\begin{figure}[ht!]
%\centering
%\includegraphics[width=\hsize]{figures/analytical_modelling.eps}
%    \caption{The best-fitting analytical model light curves of SN\,2024abfl using \cite{2016A&A...589A..53N}. 
%    The derived model parameters from the fitting are $E = 1.0025$ foe, $R=6.5\times10^{13}$ cm,and $M_\text{ej} =12.0$ M$_\odot$.}
%        \label{analytical_modelling}
%\end{figure}

\begin{figure*}[ht!]
    \centering
        \resizebox{14cm}{21cm} % 2:3
    {\includegraphics {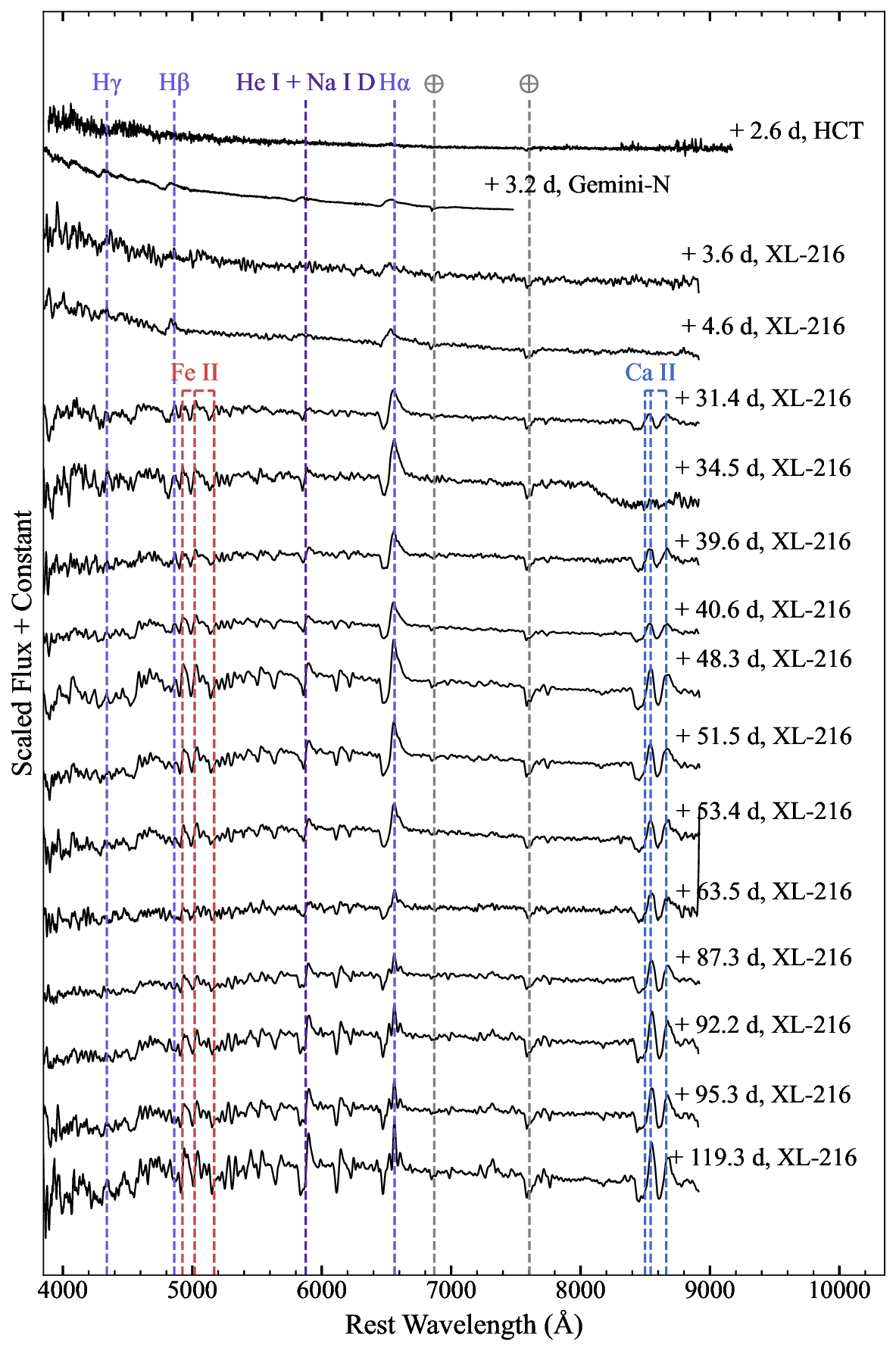}}
        \caption{Spectroscopic evolution of SN\,2024abfl.
        The displayed spectra have not been corrected for interstellar reddening. The redshift of the host galaxy ($z = 0.002979$, as reported by TNS) has been corrected. The phase represents days relative to the date of explosion, $t_0=60627.27$ MJD.}%Phases are marked relative to the first clear-band detection during the photospheric phase.}
        \label{early_middle_spectra}
\end{figure*}

%The mass of $^{56}$Ni and the recombination temperature are 0.019 M$_\odot$ and 7000 K, respectively.
%This model consists of a dense inner region and an extended low-mass envelope, designed to represent the pre-explosion structure of red or yellow supergiants—a compact core embedded within a tenuous, extended envelope.
%This approach employs a radiative diffusion framework that tracks the evolution of the recombination front within the expanding ejecta, originally developed by \cite{1989ApJ...340..396A} and later extended in semi-analytical models \citep{2014A&A...571A..77N}. 
%The ejecta are assumed to expand homologously and spherically, with a constant Thomson scattering opacity assigned to each chemically distinct layer. The time-dependent photon diffusion equation is then solved, including energy input from hydrogen recombination, radioactive decay of $^{56}$Ni and $^{56}$Co, and the increasing escape of gamma rays at late epochs.
%We fit this model to the blackbody-corrected bolometric light curve using a constant density profile for both the core and the shell, and derived the best-fitting parameters as shown in Table \ref{analytical_modelling_parm} and Figure \ref{analytical_mcmc_modelling}. 
%The best-fitting explosion energy, progenitor radius and ejecta mass are 1.0025 feo, $6.5\times10^{13}$ cm and 12.0 M$_\odot$, respectively. 
%The mass of $^{56}$Ni and the recombination temperature are 0.019 M$_\odot$ and 7000 K, respectively.

\begin{table}[ht!]
\caption {The best-fitting parameters for bolometric light curve of SN\,2024abfl using \cite{2014A&A...571A..77N} and \cite{2020MNRAS.496.3725J}.}
\label{analytical_modelling_parm} 
\centering
\renewcommand{\arraystretch}{2.0}
\addtolength{\tabcolsep}{-3pt}
\begin{tabular}{l c l}
\hline\hline             
Parameter & Best-fitting value & Remarks \\         % table heading
\hline                      % inserts single horizontal line
    R$_0$ ($10^{13}$cm) & $6.8_{-5.6}^{+1.8}$ & Initial radius of ejecta\\    % inserting body of the table
    M$_\text{ej}$ ($M_\odot$) & $8.3_{-2.8}^{+5.9}$ & Ejecta mass\\
    M$_\text{Ni}$ ($M_\odot$) & $0.009_{-0.009}^{+0.089}$ & Initial $^{56}$Ni mass\\
    E$_\text{kin}$ ($\mathrm{foe}$) & $0.42_{-0.41}^{+0.18}$ & Initial kinetic energy\\
    E$_\text{th}$ ($\mathrm{foe}$) & $0.03_{-0.03}^{+0.11}$ & Initial thermal energy\\
\hline
\end{tabular}
\end{table}

%\begin{table}[ht!]
%\caption {The best-fitting parameters for bolometric light curve of SN\,2024abfl using \cite{2016A&A...589A..53N}.}
%\label{analytical_modelling_parm} 
%\centering
%\renewcommand{\arraystretch}{1.5}
%\addtolength{\tabcolsep}{-3pt}
%\begin{tabular}{l c c l}
%\hline\hline             
%Parameter & Core & Shell & Remarks \\         % table heading
%\hline                      % inserts single horizontal line
%    R$_0$(cm) & $6.5\times10^{13}$ & $9.5\times10^{13}$ & Initial radius of ejecta\\    % inserting body of the table
%    M$_\text{ej}$(M$_\odot$) & 12 & 0.15 & Ejecta mass\\
%    T$_\text{rec}$(K) & 7000 & -- & Recombination temperature\\
%    M$_\text{Ni}$(M$_\odot$) & 0.019 & -- & Initial $^{56}$Ni mass\\
%    E$_\text{kin}$($foe$) & 0.8 & 0.1 & Initial kinetic energy\\
%    E$_\text{th}$($foe$) & 0.1 & 0.0025 & Initial thermal energy\\
%    $\kappa$(cm$^2$ g$^{-1}$) & 0.289 & 0.4 & Opacity\\
%    A$_\text{g}$(day$^2$) & $3\times10^{6}$ & $1\times10^{10}$ & Gamma-ray leakage\\
%\hline
%\end{tabular}
%\end{table}
%\subsection{Color Evolution}
%\dots
%\subsection{\textcolor{red}{\textbf{$^{56}$Ni mass from tail luminosity}}}
%\textcolor{red}{\textbf{\dots}}

%%%%%%%%%%%%%%%%%%%%%%%%%%%%%%%%%%%%%%%%%%%%%%%%%%%%%%%%%%%%%%

\begin{figure*}[ht!]
    \centering
        \resizebox{14cm}{9.3cm} % 3:2
    {\includegraphics {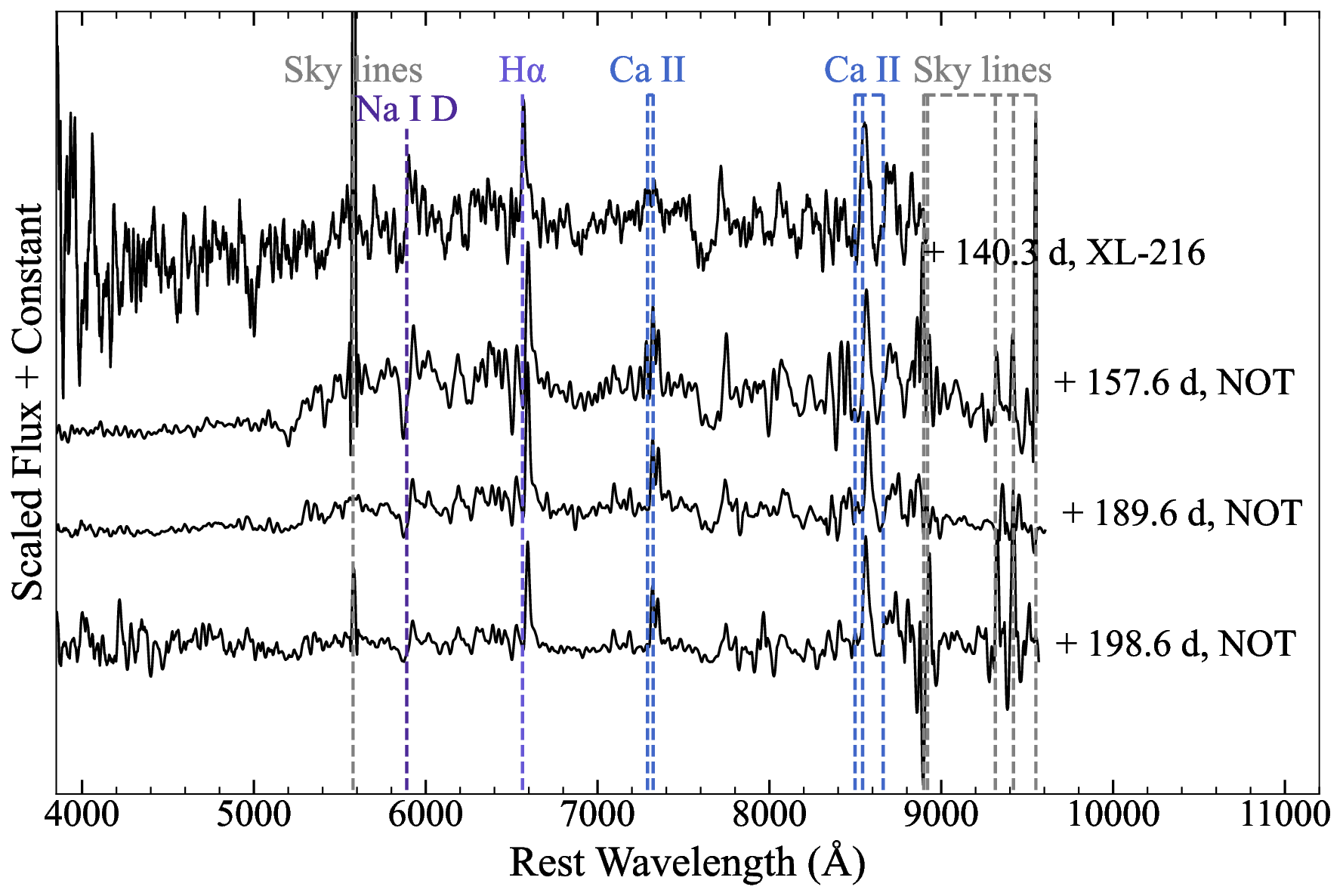}}
        \caption{Same as Figure~\ref{early_middle_spectra} but for later epochs. }
        \label{middle_late_spectra}
\end{figure*}

\begin{figure}[ht!]
\centering
\includegraphics[width=\hsize]{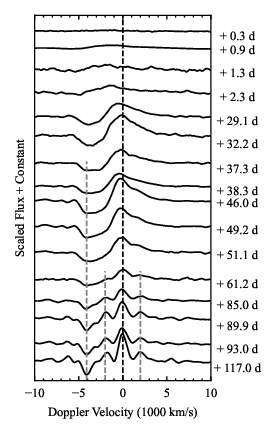}
    \caption{Evolution of H$\alpha$ line profiles during the photospheric phase.}
        \label{H_alpha_spectra}
\end{figure}

\section{Spectroscopic Analysis}\label{spec}

%\textcolor{red}{ZY: It may also be valuable to include a subsection in Section 5 comparing the observed spectra with the spectral models by Luc Dessart, similar to Fig. 10 in https://arxiv.org/abs/2512.23267v1. These models are currently among the most advanced available for Type IIP supernovae.}

%We triggered the spectroscopic monitoring of SN\,2024abfl from $t=1.3$ days to $t=196.3$ days. 
The spectral sequence of SN\,2024abfl is shown in Figure~\ref{early_middle_spectra} and~\ref{middle_late_spectra}.
From $t=2.6$ to 4.6 days, the spectra exhibit a blue continuum as the primary component, with several broad P-Cygni profiles evident atop it. 
The most prominent are the H$\alpha$, H$\beta$, H$\gamma$ and He \textsc{i} $\lambda$5876 lines.  
As the ejecta expands and cools, the continuum becomes increasingly red out to $t=31.4$ days. 
From $t=31.4$ to 119.3 days (during the plateau phase), 
%the SN continues to expand and cool, leading to reduced opacity and a progressively redder continuum. 
the H$\alpha$ emission component becomes increasingly prominent and metal lines begin to emerge, 
such as Fe \textsc{ii} ($\uplambda\uplambda$ 5169, 5018, 4924) and Ca \textsc{ii} infrared (IR) triplet ($\uplambda\uplambda$ 8498, 8542, 8662). % becomes visible in the 29.1 d spectrum and subsequently grow more prominent. 
During the nebular phase, the decay of $^{56}$Ni becomes the primary power source.
As the ejecta become optically thin, the absorption components of the P-Cygni profiles disappear. 
The spectra at $t=140.3$ days and onwards, shown in Figure~\ref{middle_late_spectra}, are dominated by H$\alpha$, Na I D, and the Ca II IR triplet emission, with the H$\alpha$ line remaining the most prominent feature.

%Figure \ref{early_middle_spectra} presents the spectral sequence of SN\,2024abfl, spanning epochs from shortly after its discovery through the plateau phase. 
%And the spectra obtained during the nebular phase is shown in Figure \ref{middle_late_spectra}.

%\subsection{Evolution of the line profiles}

%\begin{figure}[ht!]
%\centering
%\includegraphics[width=\hsize]{figures/velocity.eps}
%    \caption{Line velocity evolution of H$\alpha$, Fe II, and Ca II lines. The velocities are estimated from the absorption minima.}
%        \label{velocity}
%\end{figure}

\subsection{Kinematics}
Figure~\ref{H_alpha_spectra} shows the evolution of H$\alpha$ line profile from $t=2.6$ to 119.3 days (during the plateau phase). 
The data show a decline in expansion velocity due to the recession of the photosphere in the SN ejecta with lower velocities in the interior. 
%From $t=0.3$ to 32.2 days, H$\alpha$ P-Cygni profiles emerge in the SN spectra with a decreasing velocity. 
The velocity of H$\alpha$ absorption dip decreased from $\sim3800$ km s$^{-1}$ at $t=31.4$ days to $\sim3400$ km s$^{-1}$ at $t=34.5$ days, and further declined to $\sim2800$ km s$^{-1}$ at $t=39.6$ days.
These velocities are much lower than normal SNe~IIP at similar epochs \citep[$>4500$ km s$^{-1}$ at $<$ 50 days;][]{2003ApJ...582..905H}.
%\textcolor{red}{velocity, lower than normal IIP}
From $t=39.6$ days, a high-velocity absorption feature emerges at $-4200$ km s$^{-1}$. 
This is possibly due to a plume of matter in the inner ejecta moving toward us at a higher velocity. 
In addition, two emission features show up at $\pm2000$ km s$^{-1}$ from $t=63.5$ days. 
It is possible that the SN ejecta is colliding with two clumps of CSM, which were ejected by the progenitor before its explosion. 
\cite{2026arXiv260102638G} also found these features and explained it as the interaction with CSM. 

\subsection{Nebular phase}

At the nebular phase, the [O\,I] $\uplambda\uplambda$ 6300, 6364 lines primarily originate from the cooling radiation of the oxygen-rich zone; their intensity is directly related to the oxygen mass, serving as an important tracer for inferring the progenitor star mass. Meanwhile, the [Ca\,II] $\uplambda\uplambda$ 7291, 7323 lines mainly arise from the primordial calcium (solar abundance) in the hydrogen envelope of the progenitor, reprocessing deposited energy; their intensity reflects the mass of the hydrogen-rich region and the degree of $^{56}$Ni mixing \citep{2017hsn..book.....A}.
The flux ratio [O\,I]/[Ca\,II] reflects the mass distribution structure between the stellar core and the envelope, which is determined by the initial mass of the star. Therefore, measuring this flux ratio in nebular phase spectra allows for the estimation of the progenitor star's mass.

In our observations, the forbidden calcium emission lines [Ca\,II] $\uplambda\uplambda$ 7291,7324 are clearly visible, whereas the [O\,I] $\uplambda\uplambda$ 6300,6364 lines remain buried in the noise. 
We measured the flux of the [Ca\,II] line and estimated the upper limit of [O\,I] line flux.
We fitted the [Ca\,II] feature with a Gaussian profile, assumed that the [O\,I] lines share the same width, and adopted 3 times the continuum noise near [O\,I] as the upper limit of the line height.
% the local continuum level near [O\,I] plus $3\sigma$ as an upper limit on their line depth. 
%\textcolor{red}{Li Wenxiong: Provide the calculation methodology. Clearly detail the derivation of the upper limit (window, noise...). What sigma level does the limit correspond to?}
At $t=196.3$ days, the flux ratio [O\,I]/[Ca\,II] is smaller than 0.75. 
By comparing with the flux ratios obtained from the modeled spectra with different progenitor masses at a similar epoch \citep{2012A&A...546A..28J,2014MNRAS.439.3694J,2024A&A...687L..20F}, we estimate the progenitor mass to be $\le15$ $M_\odot$. 
This is consistent with the result of \cite{2025ApJ...982L..55L} based on the detection of the progenitor and \cite{2026arXiv260102638G} derived form the nebular spectra (9--12 $M_\odot$). 
%\textcolor{orange}{The results for 138.0, 155.3, 187.3 and 196.3 days are 1.79, 1.00, 0.90 and 0.75, respectively.
%By comparison with the ratios of 0.35, 0.60, 1.18, and 1.74 derived from modeled spectra corresponding to progenitor masses of 12, 15, 19, and 25 M$_\odot$, respectively \citep{2012A&A...546A..28J,2014MNRAS.439.3694J,2024A&A...687L..20F}, our result indicates that the upper limit of the ZAMS progenitor mass greater than 
%15 M$_\odot$, while \cite{2025ApJ...982L..55L} estimated that the initial mass of the RSG progenitor is in a range of $9-12$ M$_\odot$. 
%This discrepancy likely stems from the non-detection of the [O\,I] lines in our nebular spectrum, for which only upper limits on the flux can be derived. Consequently, the resulting constraint on the progenitor mass is weaker, allowing for a broader mass range.
%}
%0.35 0.6 (0.75)(0.90)(1.00) 1.18 1.74 (1.79)

%12 ... 15 \dots\dots\dots\dots\dots\dots 19 \dots 25

%\textcolor{red}{\textbf{the mass of progenitor \dots 
%\citep{2025ApJ...982L..55L}: the SN 2024abfl had an RSG progenitor with initial mass of $9-12$ M$_\odot$.}}
%The results are shown in Figure \ref{Ca_II_O_I_ratio}.
%the flux ratio [O\,I]/[Ca\,II]:
%\begin{enumerate}
%    \item 12 M$_\odot$: 0.35
%    \item 15 M$_\odot$: 0.6
%    \item 19 M$_\odot$: 1.18
%    \item 25 M$_\odot$: 1.74
%\end{enumerate}

\section{Summary and Discussion}\label{summary}

%\textcolor{red}{CESAR: clarify t again: most people read the abstract and conclusion first}

We present photometric and spectroscopic observations and analysis of SN\,2024abfl, a SN\,IIP in the nearby galaxy NGC\,2146.
It has a relatively long plateau phase of $\sim126.5$ days, suggesting a thick hydrogen envelope. 
The absolute magnitude of the plateau is $M_V$ $\sim-15$ mag, which is much lower than those of normal SNe~IIP. 
By fitting a semi-analytical model to the bolometric light curves, we estimated a 
$^{56}$Ni mass of $\sim0.009$ $M_\odot$, 
an initial kinetic energy of $\sim0.42$ foe, 
an initial thermal energy of $\sim0.03$ foe, 
a progenitor radius of $\sim6.8\times10^{13}$ cm and an ejecta mass of $\sim8.3$ $M_\odot$. 
The $^{56}$Ni mass is consistent with the value derived independently from the tail bolometric luminosity.%, which is $\sim0.009$ $M_\odot$. 

The spectra of SN\,2024abfl are characterized by an early blue continuum that turns red over time, several broad P-Cygni features from H$\alpha$, Ca \textsc{ii}, Fe \textsc{ii} and Na \textsc{i} D during the photospheric phase and low velocity ($\sim3800$ km s$^{-1}$ at $t=31.4$ days). 
From $t=37.3$ days, a high-velocity H$\alpha$ absorption feature emerges, possibly due to a plume of matter in the inner ejecta moving toward us at a higher velocity. 
From $t=61.2$ days, two emission features show up at $\pm2000$ km s$^{-1}$, possibly caused by CSM interaction. 
The SN enters the nebular phase from $t=138.0$ days and we estimate the progenitor mass to be $\le15M_\odot$ by the flux ratio of [O\,I]/[Ca\,II]. This is consistent with the progenitor mass inferred from the direct deteciton. 

\cite{2026arXiv260102638G} reported a slightly fainter peak absolute magnitude of SN\,2024abfl in the $V$ band as $-14.9$ mag, which is likely due to their adoption of a lower extinction value ($E(B-V)_\mathrm{tot}=0.28\pm0.11$) compared to our estimates. 
The low values for the $^{56}$Ni mass, ejecta mass, and explosion energy estimated by \cite{2026arXiv260102638G} and \cite{2025arXiv250620068D} are consistent with our values within uncertainties, supporting the classification of SN\,2024abfl as a LLSN.

Compared with the typical range for SNe~IIP ($^{56}$Ni$~$mass: 0.001--0.26 $M_\odot$; explosion energy: 0.1--5.5 foe; ejecta mass: 5.4--24.8 $M_\odot$), SN\,2024abfl has a lower $^{56}$Ni$~$mass, lower explosion energy, and lower ejecta mass. 
These parameters fall within the range of LLSNe~IIP ($^{56}$Ni$~$mass: 0.001--0.025 $M_\odot$; explosion enery: 0.1--0.28 foe; progenitor mass: 8--12 $M_\odot$) \citep{10.1111/j.1365-2966.2004.07173.x,2025PASP..137d4203D,2025arXiv250620068D}.
The low progenitor mass \citep[9--12 $M_\odot$;][]{2025ApJ...982L..55L,2026arXiv260102638G} once again provides evidence that LLSNe~IIP originate from low-mass RSG progenitor stars.

%\textcolor{red}{velocity, mass ... lower than IIP, consistent with LLSNe~IIP~IIP; progenitor mass, once again prove low mass; detail value of IIP, LLSNe~IIP~IIP, compare with our value}

%\textcolor{red}{
%    \begin{enumerate}
%        \item $^{56}$Ni mass: $0.007_{-0.007}^{+0.042}$
%        \begin{enumerate}
%            \item 0.010$\pm$0.001\citep{2026arXiv260102638G}
%            \item $0.02_{-0.017}^{+0.015}$\citep{2025arXiv250620068D}
%            \item $0.0038_{-0.0002}^{+0.0003}$\citep{2025arXiv250620068D}
%        \end{enumerate}
%        \item $M_\mathrm{ej}$: $6.3_{-2.1}^{+1.0}$
%        \begin{enumerate}
%            \item $5.55_{-0.32}^{+0.28}$\citep{2025arXiv250620068D}
%        \end{enumerate}
%        \item $E_\mathrm{kin}$: $0.22_{-0.09}^{+0.20}$
%        \begin{enumerate}
%            \item $0.21_{-0.01}^{+0.01}$\citep{2025arXiv250620068D}
%        \end{enumerate}
%        \item $E_\mathrm{tot}$: $0.23_{-0.1}^{+0.27}$
%        \begin{enumerate}
%            \item 0.12\cite{2026arXiv260102638G}
%            \item $0.1_{-0.00}^{+0.04}$\citep{2025arXiv250620068D}
%        \end{enumerate}
%    \end{enumerate}
%}

\section*{Acknowledgements}
    ZF is supported by the Strategic Priority Research Program of the Chinese Academy of Sciences, Grant No. XDB0550100. 
    NCS is funded by the Strategic Priority Research Program of the Chinese Academy of Sciences Grant No.XDB0550300, 
    the National Natural Science Foundation of China Grants No.12303051 and No.12261141690, and the China Manned Space Program No.CMS-CSST-2025-A14. 
    We acknowledge the support by the National Astronomical Research Institute of Thailand under the project number TRTC11A\_004, TRTToO\_2024002, TRTToO\_2024004, TRTToO\_2024005, TRTToO\_2024006, TRTC12A\_003, TRTC12B\_003 and TRTC12C\_003.
    JL is supported by the National Natural Science Foundation of China (NSFC; grant No. 12273027). 
    ZG is supported by the China-Chile Joint Research Fund (CCJRF No.2301) and the Chinese Academy of Sciences South America Center for Astronomy (CASSACA) Key Research Project E52H540301. This work was supported by the National Research and Development Agency (ANID) through the FONDECYT Iniciación project Grant No. 11260176. 
    FP acknowledges support from the MICINN under grant numbers PID2022-141915NB-C21.
    ZXN is funded by the NSFC Grant No. 12303039.

    We thank the staff at all participating observatories, including the Xinglong-35cm, 60-cm and 2.16-m telescopes, the Liverpool Telescope, the Thai Robotic Telescope network, and the Nordic Optical Telescope, for their support during our observing campaigns.

    We acknowledge the use of data and services provided by ATLAS, ZTF, Swift/UVOT and the Transient Name Server (TNS). This research has made use of NASA's Astrophysics Data System (ADS).
    The ZTF forced-photometry service was funded under the Heising-Simons Foundation grant \#12540303 (PI: Graham).
    We also acknowledge the use of open-source software packages including ASTROPY, AUTOPHOT, and IRAF. 
    We also acknowledge the use of open-source software packages including astropy \citep{2018AJ....156..123A,2022ApJ...935..167A}, AutoPhot \citep{2022A&A...667A..62B}, IRAF \citep{1986SPIE..627..733T} and SuperBol \citep{2018RNAAS...2..230N}.

    \textit{Facilities}: LT, NOT, TRT, XL-35, XL-60, XL-216

    \textit{Software}: astropy \citep{2018AJ....156..123A,2022ApJ...935..167A}, AutoPhot \citep{2022A&A...667A..62B}, IRAF \citep{1986SPIE..627..733T}, SuperBol \citep{2018RNAAS...2..230N}
%\end{acknowledgements}

%%%%%%%%%%%%%%%%%%%%%%%%%%%%%%%%%%%%%%%%%%%%%%%%%%%%%%%%%%%%%%
\bibliographystyle{aa}
\bibliography{sn2024abfl}

\end{document}